\documentclass[manuscript,screen]{acmart}
\AtBeginDocument{%
  \providecommand\BibTeX{{%
    \normalfont B\kern-0.5em{\scshape i\kern-0.25em b}\kern-0.8em\TeX}}}

\setcopyright{acmcopyright}
\copyrightyear{2018}
\acmYear{2018}
\acmDOI{XXXXXXX.XXXXXXX}

\acmConference[Conference acronym 'XX]{Make sure to enter the correct
  conference title from your rights confirmation emai}{June 03--05,
  2018}{Anonymous}
\acmBooktitle{Woodstock '18: ACM Symposium on Neural Gaze Detection,
 June 03--05, 2018, Woodstock, NY} 
\acmPrice{15.00}
\acmISBN{978-1-4503-XXXX-X/18/06}

\begin{document}

\title{Studying Artist Sentiments around AI-generated Artwork}

\author{Safinah Ali}
\email{safinah@media.mit.edu}
\affiliation{%
  \institution{Massachusetts Institute of Technology}
  \city{Cambridge}
  \state{MA}
  \country{USA}
  \postcode{02143}
}
\author{Cynthia Breazeal}
\email{cynthiab@media.mit.edu}
\affiliation{%
  \institution{Massachusetts Institute of Technology}
  \city{Cambridge}
  \state{MA}
  \country{USA}
  \postcode{02143}
}

\renewcommand{\shortauthors}{Ali \& Breazeal}

\begin{abstract}
 Art created using generated Artificial Intelligence (AI) has taken the world by storm and generated excitement for many digital creators and technologists. However, the reception and reaction from artists have been mixed. Concerns about plagiarizing their artworks and styles for datasets and uncertainty around the future of digital art sparked movements in artist communities shunning the use of AI for generating art and protecting artists’ rights. Collaborating with these tools for novel creative use cases also sparked hope from some creators. Artists are an integral stakeholder in the rapidly evolving digital creativity industry and understanding their concerns and hopes inform responsible development and use of creativity support tools. In this work, we study artists’ sentiments about AI-generated art. We interviewed 7 artists and analyzed public posts from artists on social media platforms Reddit, Twitter and Artstation. We report artists’ main concerns and hopes around AI-generated artwork, informing a way forward for inclusive development of these tools. 
\end{abstract}



\begin{teaserfigure}
  \includegraphics[width=\textwidth]{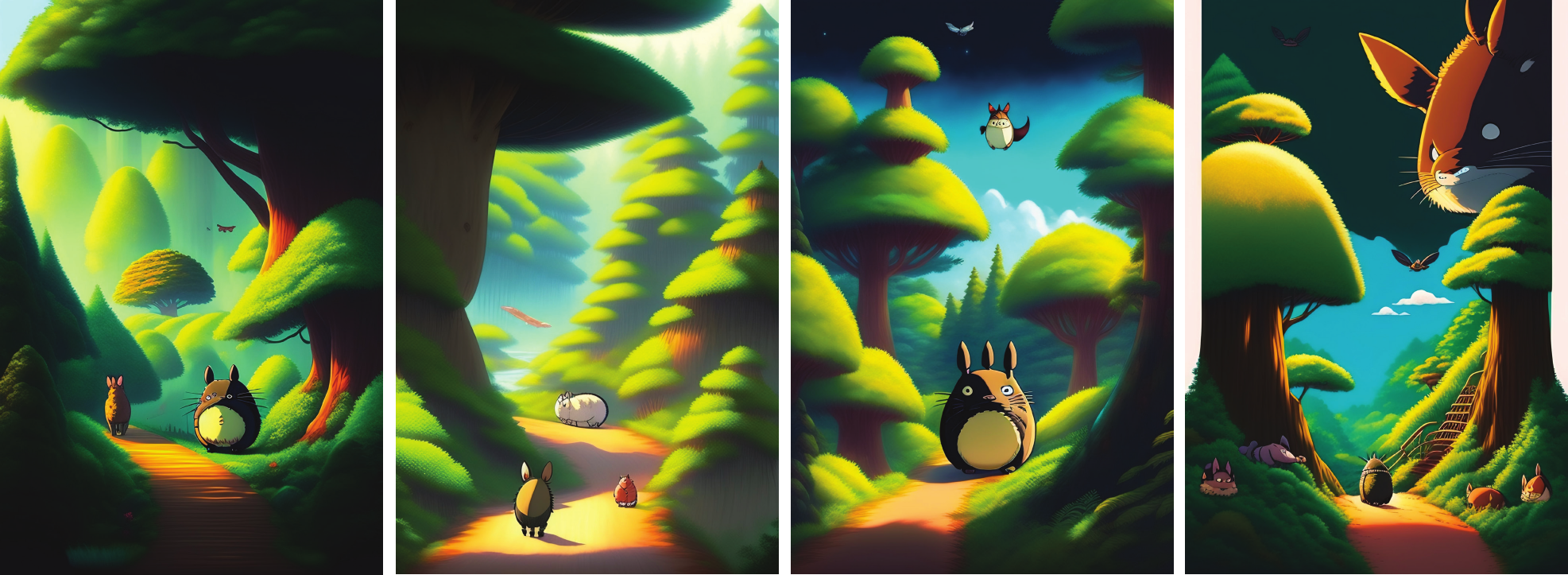}
  \caption{Images generated using the prompt "Studio Ghibli style characters in the forest" in Stable Diffusion. Source: Lexica.art}
  \Description{Images generated using the prompt "Studio Ghibli style characters in the forest}
  \label{fig:teaser}
\end{teaserfigure}

\maketitle

\section{Introduction}
Text-to-image generation algorithms, a new way to generate digital visual art has taken the internet by a storm. Surreal, beautiful, vivid and sometimes funny imagery emerged from these AI algorithms. What also emerged was a divided opinion from the visual artist community. Some welcomed new modalities of creation while others opposed these tools, giving rise to the “Say NO to AI Art” movement. Visual artists are a key stakeholder in the emerging field of generative art. Their work and livelihood can be affected by the development and proliferation of AI-assisted generation tools. However, given the rapid nature of how these tools have been evolving, artists’ views on how generative AI algorithms impact their lives has not been systematically studied. 

AI algorithms and tools such as Stable Diffusion, Dall-E and Midjourney present the ability to generate visual images from natural language prompts. These algorithms are trained on large datasets of images paired with natural language descriptions (often captions or alt. Text accompanying images on the internet) and learn to generate novel realistic images matching a new language prompt. When trained on datasets from specific art styles, for instance, on art with Pointillism, it learns to generate art with tiny dots of pure colors. Artists have undoubtedly been pivotal in the development of these algorithms, especially to make them aesthetically pleasing, since it is their art that served as a part of training datasets that powered these algorithms, and made them especially aesthetically pleasing. However, artists have since been left out of the conversation of what has primarily become a tech-driven field. Artists have taken to various social media platforms and art sharing websites to express their views and discuss implications of AI art with other artists. In this work, we conduct a systematic review of visual artists’ perceptions of AI-generated art. We aim to understand their initial feelings, concerns, hopes and opinions on policy regulating AI-generated art. 

We conducted interviews with seven visual artists and collected data from publicly available discussion about AI-generated art on social media websites to form insights into artists’ perceptions of AI-generated art. Our research goal was to understand artists’ hopes around the use of AI-powered art generation tools, their concerns about the proliferation of these tools, how they see these tools influencing their work, and their opinions on what policy surrounding the use of these tools must be. We conducted a thematic analysis with an inductive coding approach on the data collected. Our research questions for this work were: 

\begin{itemize}
  \item \textbf{RQ1:} What are artists’ hopes about the development and use of generative AI tools? 
  \item \textbf{RQ2:} What are artists’ concerns about the development and use of generative AI tools? 
  \item \textbf{RQ3:} How do generative AI tools influence artists’ work and life? 
  \item \textbf{RQ4:} What are artists’ opinions about policy surrounding the development of generative AI tools? 
\end{itemize}

Artists are primary stakeholders in creative fields, and hence understanding their needs and concerns informs ethical and responsible development of creativity support tools. Our findings inform AI developers for responsible and inclusive design of automated generative tools, artists for being better equipped to understand the potential impact of AI tools on their work and understanding their peers’ perceptions of these tools, and policy makers for designing regulations around the development and use of these tools. 

\section{Background}

\subsection{AI-generated Art }

Generative AI algorithms are type of AI algorithms that are used to generate novel data samples that are classified as belonging to the training data samples. Generative AI algorithms are used to create a variety of media and have found applications in art, imaging, engineering, protein folding and modeling. Generative AI algorithms first came into being with the advent of Generative Adversarial Networks (or GANs) in 2014~\cite{goodfellow2020generative}. GANs made use of two neural networks: Generator network and Discriminator network that work in opposition to learn to generate novel data with the same statistics as the training dataset. GANs were used in several creative applications such as creating stylized images from sketches~\cite{karras2020analyzing}, generating artwork from photographs~\cite{isola2017image}), or even creating fake faces~\cite{wang}. Since the advent of GANs, generative algorithms have come a long way and the quality of the generated media have become increasingly detailed and realistic. The availability of large datasets such as LAION-400M~\cite{schuhmann2021laion} have played a major role in the advancement of these algorithms. 

A revolution in generative AI algorithms occurred when generative models were coupled with Large Language Models (LLM) such as CLIP~\cite{radford2021learning} to generate images from natural language prompts. These came to be known as text-to-image generation algorithms. This gave rise to tools such as Stable Diffusion, Dall-E or Midjourney. These models couple a generative model (typically a Diffusion model~\cite{rombach2022high}) that learns to generate high resolution images using a fixed forward noising and reverse denoising of images, with a natural language model (typically encoder of CLIP) which has been trained on millions of image-text pairings and determines the match between the generated image to a given prompt. Adding artistically relevant terminology to prompts enabled the creation of aesthetically pleasing artwork. Adding artist names, art styles or art studio names lead to the generation of the corresponding style of images. For example, all images in Figure~\ref{fig:teaser} were generated with “Studio Ghibli” in the prompt.

\subsection{Reception of AI-generated Art}
\begin{figure}[h]
  \centering
  \includegraphics[width=0.5\linewidth]{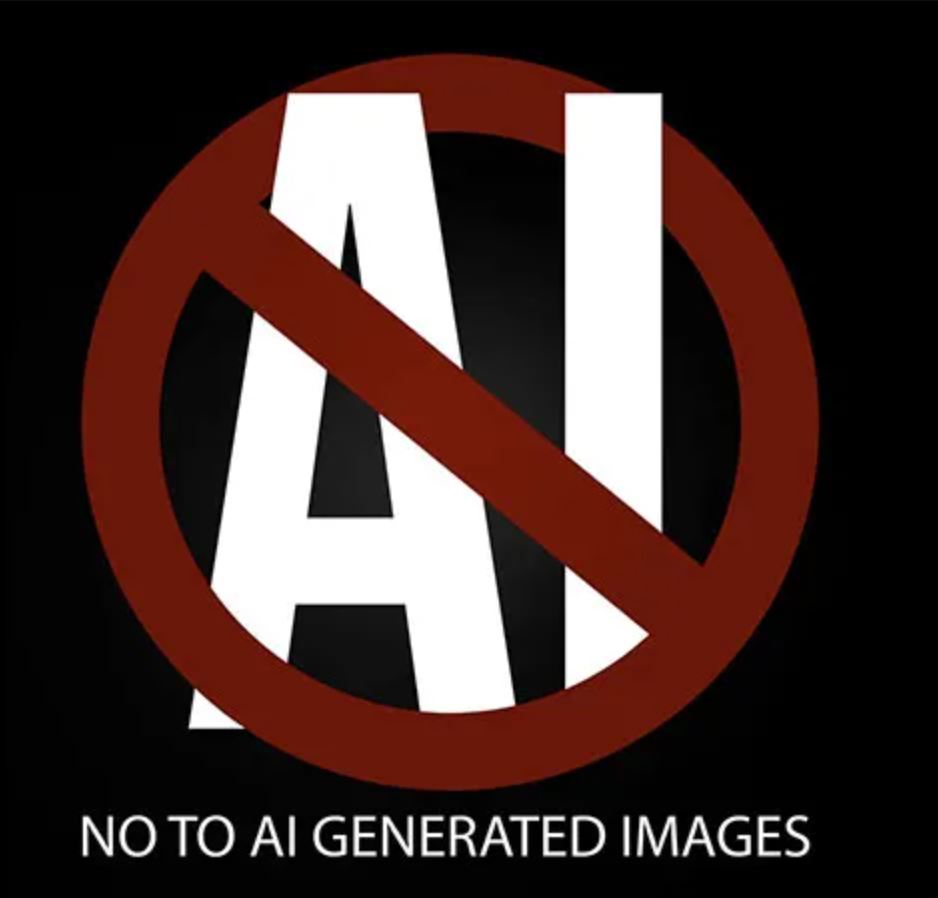}
  \caption{Poster shared by artists on Artstation to protest AI Art}
  \Description{Poster shared by artists on Artstation to protest AI Art}
  \label{fig:poster}
\end{figure}

Unlike previous versions of generative AI models, recent tools such as Stable Diffusion or Midjourney are also openly available for anyone to use and do not require a sophisticated computational or Graphical Processing Unity (GPU) setup. This rapidly augmented the generation and spread of AI-generated art on the internet and art showcasing platforms. The reception of AI-generated art has, however, been mixed. While some are amazed at the abilities of this rapidly evolving technology, others view it with skepticism of not being “real” art. Users of the tools have expressed excitement at the creative possibilities, especially the ability to image anything and generate its visuals in a few seconds. Users have also used these tools to generate art in the style of specific artists. For instance, Other users have categorized the use of these models as plagiarism, especially when they used datasets from artists without their consent. Some users have expressed their concerns about how these tools would impact creative industries and individuals. Some artists have also participated in online protest movements against AI-generated art, such as the “No to AI-generated Images” movement, where they unfollowed accounts posting AI-generated art, and shared posters rejecting AI Art (Figure~\ref{fig:poster}). It is clear from a myriad of responses that the understanding of the impact of these tools is complex and these systems must not be viewed simply as technical systems, but socio-technical systems~\cite{emery2016characteristics} with lasting social impact, especially for creators. While there is a lot of rich data about artists’ reactions to these tools, in our knowledge, there is no systematic study of artists’ perception of the tools, which we aim to do with this work.

\section{Methods}
\subsection{Data Collection}
\subsubsection{Social Media Platforms}
We collected artists’ public comments made on social media platforms Reddit, Twitter and Artstation.com. Reddit is a news aggregation, content rating, and discussion website. Topic-wise organization in the form of subreddits allows for rich discussions pertaining to one topic. Reddit also houses communities, or individuals with shared interests, often commenting on topics of shared interests. We collected data from subreddits used by visual artists, AI artists as well as creators of generative AI models. Twitter is a microblogging and social networking website. Users often take to Twitter to demonstrate support towards or protest against emerging services. Comment responses and quote tweeting enable users to agree or disagree to the original tweet sparking broader community discussion. Artstation.com is an art sharing website where digital art creators commonly share their art. Comments on Artstation.com is a space for users to discuss the art and the tools that the artist used. Artstation.com has also been used as a community protest tool, for instance, users posting posters against war during the Ukraine-Russia conflict in 2022, or users posting graphic art to protest AI art. These platforms were chosen because these were spaces where rich discussions about AI art, its possibilities, its repercussions and how it influences artists occurred and comments were available to the public. These platforms were also commonly used by artists to share their work, and host community discussions surrounding their work and creative tools. There are other platforms where such discussions took place, such as Deviantart.com, Instagram, and Facebook. However, we did not use data from these platforms because either the platform did not provide publicly accessible dataset for aggregation, or there wasn’t enough relevant data found pertaining to our research questions. 

We collected our data using hashtag and keyword searches. We used the keywords “AI art”, “Generative Art”, “Stable Diffusion”, “Midjourney”, “AI-generated”, “Art model” for our initial search. After initial data collection and identifying keyword patterns, we also included “AI plagiarism”, “DallE art”, “Support AI Art”, “Say No to AI art”, and “Ban AI Art”. Data was collected from posts, comments and replies on all platforms. For this research, we only included data pertaining to our research questions and only from users who identified as artists, either in their post or their profile. 

\subsubsection{Artist Interviews}

We also interviewed seven visual artists about their perceptions about AI-generated art. We recruited by contacting artists who share their digital artwork through their social profiles. 50 artists were contacted and we attempted to balance across art medium, age groups, and gender. We received positive interest from 11 artists, and were able to schedule interviews with seven artists (Age range: 22-43, M = 30.429, S.D. = 6.651; Gender: 3F, 4M). All participants consented to the interviews being recorded and used for research purposes (including for publication and presentation materials). The following questions were asked during the interview: 

\begin{itemize}
  \item How would you best describe your art form?
  \item What art tools or platforms do you use to create / share / sell your art? 
  \item Are you familiar with AI-generated art? Can you describe your experience with it?
  \item What were your initial feelings about AI-generated art?
  \item What are some of your concerns about AI-generated art?
  \item What are some of your hopes about AI-generated art?
  \item Do you support the advancement of AI-generated art? Can you explain why?
  \item Do you think there should be policy regulating the generation and use of AI-generated art?
\end{itemize}

Artists were given the option to skip the questions they did not want to answer or did not have enough information to answer. Artists were also told that they can choose to terminate their participation at any time during the interview. All interviews were conducted on the video conferencing tool Zoom and participants could share screen or send materials via chat if they wanted to elaborate on a topic. 

\subsection{Data Analysis}
We used a reflexive thematic analysis approach~\cite{braun2019reflecting} with an inductive coding practice for data analysis. All information collected from the social media platforms and artist interviews were listed as ideas or insights in a sheet. An insight consisted of a sentence or group of sentences pointing to the same topic that the artist spoke or wrote about. Two coders then generated succinct labels or codes that capture important features of the data, especially in relevance to the research questions. They used an inductive approach while coding the data. This meant that the coders did not use predetermined codes but formed the codes as they analyzed the data. 

For instance, the interview insight, “They [technologists] complain that artists are gatekeeping art, but it’s not that - artists are angry that their work has been blatantly stolen from them and companies and platforms have failed them.” was coded as “plagiarism of artists’ dataset” and “artists betrayed by platforms”. 

Coders discussed differences in assigned codes and collaborated to generate unified codes for all insights. In accordance with the reflexive thematic analysis process, we first identified significant broad patterns in these codes to generate initial themes. For instance, “resentment over plagiarized datasets” was a broad initial theme that emerged. Codes relevant to initial themes were collated to review the viability of each theme. Post collation, we reviewed the themes against the coded data to split, combine or discard themes. Themes with insufficient supporting data or that were irrelevant to the research questions were discarded. We then developed a detailed analysis of each theme, determining the narrative of each given the coded dataset. In the next section, we reported the final reviewed themes from this dataset.

\section{Findings}

We collected 2411 posts and comments: 1608 from Twitter, 592 from Reddit and 211 from artstation.com. We conducted one-hour interviews with seven artists. Artists were in the age range 22 to 43 (M = 30.429, S.D. = 6.651). Three artists identified as female and four identified as male. All artists described their artform as a type of visual art, including, digital painting, character design, acrylic painting, drawing, fashion design, animation, visual effects, illustration, game design, computational art, and paper sketching. In this section, we present the themes identified while conducting a thematic analysis of data collected from social media platforms and artist interviews, in lieu of the research questions of this study. 

\subsection{RQ1: What are artists’ hopes about the development and use of generative AI tools?}
\textbf{Generative AI tools open new possibilities of visualizing imaginations.}
Artists recognized new possibilities in visual art made possible through the use of generative AI. In our data codes, the word “imagin(e/ation)” occurred 577 times. One interviewee said, “I use it as a way to create these imaginary fantastical characters.” Another interviewee described their use of Midjourney for visualizing imaginations - “for me, it is really a play tool where I could put a lot of my fantastical imagination to context and I found myself obsessively creating every time I had a little free time”. Another interviewee commented, “it fascinates me what machines interpret words as”. Yet another interviewee said, “I see it as a great tool for visualization, particularly a space where you can push your ideation and imagination, and with the help of the tool, find means to use this in the context of other artistic expressions.”. A creator on Reddit shared their excitement about visualizing their imaginations of scenarios that they dream of but were unlikely to exist in our current reality, such as, “a mother breasfteeding in a spaceship.” Several data codes pointed to generative tools enabling the visualization of imaginary or unlikely scenarios, or recreating dreams. 

\textbf{Generative AI tools can be used as collaborative tools in creative fields.}
When asked about how they used generative AI tools or their hopes about the use of these tools, most artists described how it can be used as a collaborative tool. One interviewee said, “I also would like to see how artists use it as a collaborative tool for idea generation - how are they tweaking on their generated media to get to what they intend?” Artists commented on how they use it as an intermediate part of their creation workflows - “I often make variations of my own art. I have generated game scenes, ads, characters, settings. I have taken characters and animated them in my own tools.” Another artist commented, “I like it as a brainstorming tool. I use it as a way to create these imaginary fantastical characters. I make tweaks to existing characters.” A Reddit user used Midjourney + DallE + Photoshop to create characters for their game, and shared them with their community for feedback. We encountered several examples of creators trying out multiple variations of the art they generated in the first try to reach the desired image. One user created new 

\textbf{Artists are impressed by the abilities of generative AI tools.}
Five out of seven artists we interviewed as well as several accounts from social media users expressed their surprise by how good the generative AI tools are and were impressed by their abilities. One interviewee said, “I've worked with surrealism and hybrid creatures in a lot of my works, across multiple mediums and platforms, and I was fascinated at the possibilities of what I was able to create through AI”. Another artist who was otherwise skeptical about the spread of AI-generated art said, “I think the tools themselves are impressive in their ability”. An Artstation user used DallE to create new AI materials. They reflected, “I am blown away by it's abilities!
All of these base images were created by DALL-E2 AI, I then offset them in Photoshop and bought them back to DALL-E2 to paint out the seams and get it to fill in the gaps. I was astonished when this actually worked.” 

\textbf{Generative AI tools make creation more accessible.}
Three out of seven interviewees commented on how more people now have access to creating digital art. One interviewee said, “I hope that a lot more people get into making art. I hope to see people use it in interesting use cases - Like I saw someone generated book titles and see what the corresponding cover would look like.” Another interviewee said, “Lots of people will be able to create their own images and you can sort of generate these things you can only imagine.” Commenting on tools increasing accessibility to creating art, however, was not indicative of how positively artists felt about generative art. For instance, while one interviewee was excited that any person can now create in famous art styles, one artist expressed that they were “heartbroken” that anyone could now generate what they “worked so hard to learn to create”.

\begin{figure}[h]
  \centering
  \includegraphics[width=0.5\linewidth]{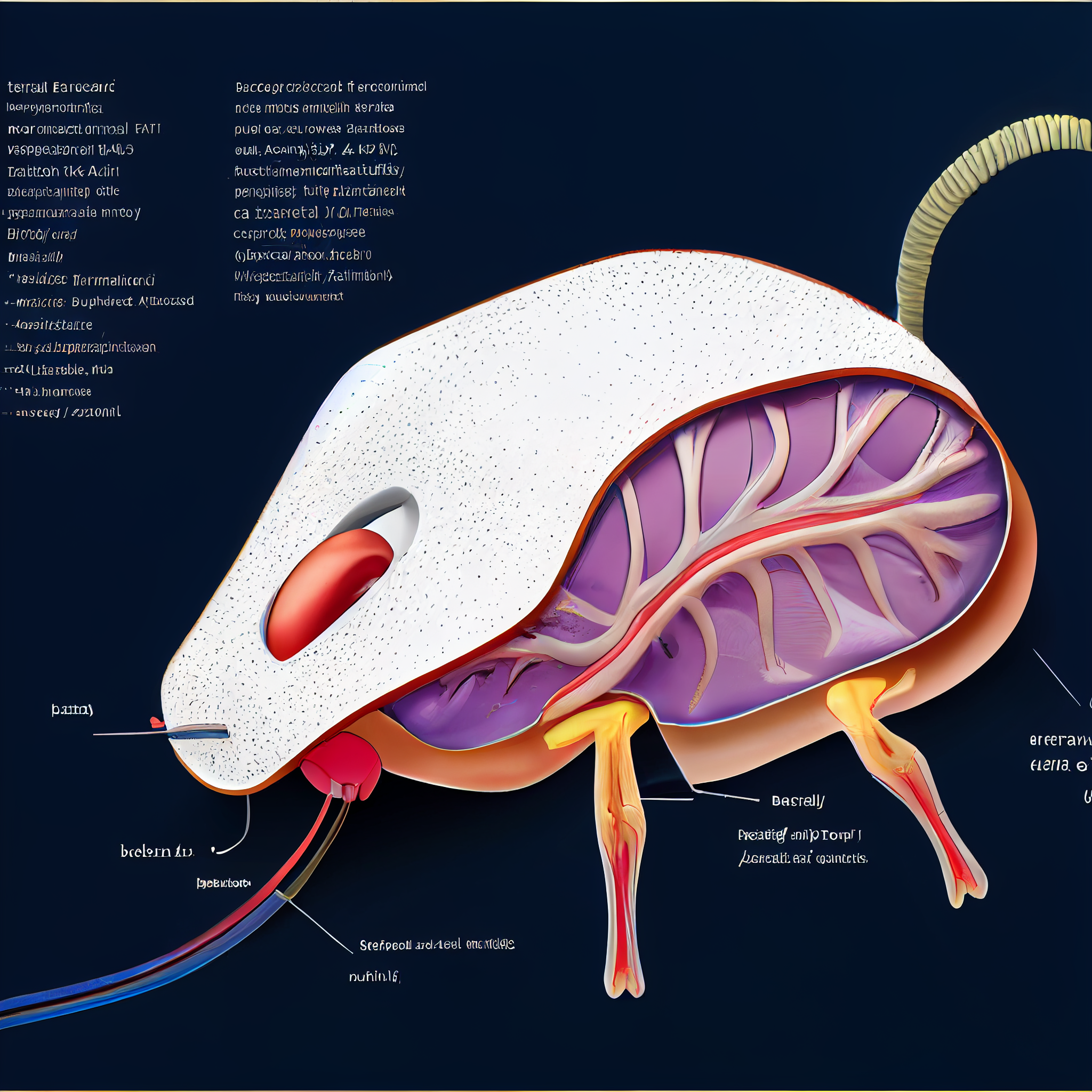}
  \caption{Image generated using the prompt “Biologically anatomical description of a computer mouse”. Source: Reddit.}
  \Description{Image generated using the prompt “Biologically anatomical description of a computer mouse”}
  \label{fig:mouse}
\end{figure}

\textbf{Imperfections of generative AI tools inspire delight.}
Several accounts of artists sharing AI-generated art with their community involved being delighted or surprised by what AI-generated - especially if it deviated from the typical human imagination of the same prompt. While describing possible use cases of generative AI, an interviewee said, “I think it can be used for humor. Most generations are quite funny or surprising to me.” One Reddit user generated an image with the prompt “Biologically anatomical description of a computer mouse”. The resulting image (Figure~\ref{fig:mouse}) was inaccurate but both the unreadable prompts and the anatomical interpretation of AI were amusing to the creator and other viewers.

\textbf{Artists acknowledge that the adoption of the tools may evolve with time.}
Another common theme that emerged was the comparison with other technologies that took time for adoption. One artist said, “I remember that Digital Art wasn't considered real art, until it was, and I hope people can see Ai Art soon enough as Art and not undermine it because it is "generated". It still needs a concept and imagination.” They further commented, “We can't and shouldn't hide from technology. It is inevitable and its perhaps best that we embrace it, adapt to it, adopt it and continue as artists to create. Photography came into being because Cameras were invented, up until then Artists physically painted photographs, and today phone companies are obsessively pushing Cameras as their primary feature of upgrade. Artists physically ruled and drew publications that went into Print, until Adobe came and revolutionized Digital Art.” Another interviewee commented, “I don’t think people who are upset about ‘oh this is not true art, etc’ are folks I empathize with - because with every technology, art has changed. This is not new, we have seen it happen with cameras, photoshop, programming, even better paints for that matter.” Another Reddit user commented on artists that are upset about AI art, “Don’t take it personally it’s a tale as old as time and we've heard it all before”, and went on to list technologies that drew opposition including photoshop, cameras and techno music. 

\subsection{RQ2: What are artists’ concerns about the development and use of generative AI tools?}
\textbf{Artists are hurt and angry about use of their datasets in AI models without their consent.}
The most common concern about generative AI tools that emerged from our dataset was about the use of artists’ datasets without their consent. The theme of plagiarism occurred in 1102 codes and was mentioned by all seven interviewees while describing their concerns. This was especially angering to artists whose art was used to train a model, and then generate images in their style. An illustration artist on Twitter expressed that she was “heartbroken” seeing users generate images in her art style in seconds, and added that she “trained thousands of hours to gain this skill, and it was stolen overnight.” Artists on social media as well as six out of seven artists we interviewed categorized this use of artists’ work in datasets without their permission as “plagiarism” or “theft”. For instance, one interviewee said, “I am against the use of artists’ art without their permission. Artists could be from somewhere else, dead or alive - it is not okay unless they explicitly permitted them.”  Artists whose style was not explicitly recreated using these models also protested in solidarity with other artists. For instance, Twitter users criticized a developer who trained and shared a model to draw in the style of late artist Kim Jung Gi as homage. One artist said, “christ this is disgusting and the audacity to ask for credit”, when the developer asked for users of the model to give credit to them. Another commenter said, “they have no shame”. Even artists who were otherwise positive or neutral about the use of AI tools for creating art were critical of using artists’ work without their permission. One interviewee said, “I am with my fellow artists in opposing AI-generated stuff and truly despise these so called artwork. I don’t think I hate the technology in particular - it’s quite intriguing, but the way these big companies and developers have gone about stealing artists’ data is quite awful.” Another artist who had created over a thousand pieces of imagery using AI said, “I understand though, that a lot of images that AI is sourcing are not ethically sourced and this remains to be addressed, at the same time, with how open source it is, and how some of these work in public domain, it is easy for others to pick your prompts and replicate them into their own, which takes away the originality of your idea and that isn't something I'm particularly a fan of.” 

\textbf{Artists feel betrayed by art sharing platforms.}
Artist recognized that it was not just the AI developers and their companies responsible for the plagiarism of their art, but also the platforms that hosted their artwork. One interviewee explained, “They complain that artists are gatekeeping art, but it’s not that - artists are angry that their work has been blatantly stolen from them and these companies and platforms have failed them.” Another interviewee said, in support of their fellow artists, “Basically, they put their art online, and sure it would have a watermark at best - but they put it on art websites and blogs - places they trust. And now all this data is mass scraped.” Similarly, another interviewee said, “I think the websites hosting art should especially be taking a strong stand against stealing this data from their sites. These sites have benefited from artists for a long time - it’s time for them to protect artists. I don’t expect tech companies to act from benevolence, but perhaps legal actions against stealing art styles or digital art without permission should be punished. I think these web platforms basically sold us after profiting from us for ages.”

\textbf{Artists are hesitant to share AI-generated art with their artist communities
.}
Through our interviews and analyzing comments on Artstation and Twitter, we observed several instances of artists afraid to post AI-generated art due to the backlash that AI-generated art receives in their communities. For instance, one artist commented, “I actually want to share some of my generated art with my communities but I am scared because the sentiment is really negative around it right now.” An Artstation user took to Reddit to seek support for the backlash they received while sharing AI-generated art on Artstation.com. 

\textbf{Artists lack trust in the developers of AI tools.}
While speaking critically of generative AI models, artists also expressed their lack of trust in the developers of these AI tools. While they hoped for their art to not be used in training these models, one artist said, “Without regulation, I don’t expect a developer to honor that.” Another interviewee commented, “I support technology in general - but I don’t trust tech experts to take into account artists and get them their  proper rights - over their work at least. I have stopped putting work online and so have many of my friends because we are scared of our work becoming parts of these datasets with zero consent.” Another interviewee commented, “I work in tech and I know about the power dynamics that exist between artists and tech folks. To be honest, they don’t care. Some might, but generally they don’t understand or care about artists. They are just excited about this new tool. And sure the new tool is great and has cool use cases, but if it’s legal, they will just keep making these tools and are not stopping just because artists are hurt.” Another interviewee said, “I just want to clarify that I am not against technology, even technology for art in general, but this sheer disregard for artists’ own work is very upsetting - especially coming from AI and big tech where they anyway don’t regard non tech persons.” Artists on social media threads were even more critical of developers of AI tools, most commonly criticizing them to be “lazy” or “stealing art”. Artists participating in the movement against AI Art also unfollowed those posting new generative models or using them to create art. One Artstation user commented, “I have already un-watched someone. They claimed they still worked on their latest post a lot in Photoshop, but you could still tell from a mile away it was AI.” Another user commented, “I was blocking every AI artist. Guess what... there's a limit to how many people you can block. And seeing how AI bros just keep sprouting up like weeds... there was and is no point.” 

\textbf{AI-generated images are not art.}
Three of the seven interviewees did not consider AI-generated images as art. One interviewee said, “I don’t think it is art. It is not original. I think it’s blatant copying of people’s style without giving them credit.” Another interviewee explained, “I think tech bros with no sense of aesthetic are randomly generating gibberish media and calling it art is what is awful. They lack the context and background to this art. Which is not to say that they shouldn’t make art - but they should be respectful of who this data is coming from and understanding what is truly aesthetic instead of feeding stuff into a machine.” Another interviewee said, “There’s no human creativity, I find it difficult to call it art.” Artists on Artstation referred to AI-generated images as “art” with quotation marks, as if with sarcasm. One Artstation user said, “This might be beautiful but I don’t think it is art, there is no human intentionality.” Another user commented, “I haven't seen any AI- generated art that had actual artistic merit.  Personally, I think what we're seeing now is more a fad that people will forget soon, rather than a new style of creating.  Someday, there may be actual AI that can create art- but that's over the horizon.” 

\subsection{RQ3: How do generative AI tools influence artists’ work and life?}
\textbf{Artists and art students are uncertain about their future.}
A common fear expressed by many artists and art students on social media communities was an uncertainty about their future career. One art student on Twitter commented, “I am in school to study Graphic Design and now I am really worried about AI taking over and if there are any prospects here.” Another Artstation user commented "And as much as I don't want to be rude, if anyone I watch starts posting AI regularly, you lost me there, I'm sorry, but I will not enable this mess any longer, staying silent IS enabling and using AI IS enabling the destruction of real art, jobs and people's livelihoods!” After plagiarism, the concern that we encountered most commonly was a loss to artists’ livelihoods. One interviewee said, “Of Course digital artists’ livelihood and value getting affected. But also, our works being stolen. I don’t know what my next steps should be - should I take down my art from everywhere? Is it too late to do that now? It even makes me demotivated to make more more art because now anyone can do it using a line. “ Another interviewee said, “If everyone could make Picasso art, how would Picasso have sold?” 

\textbf{Artists are hesitant to share their artwork online.}
In fear of their work being stolen for datasets to train AI models, artists shared that they are now hesitant to share their work online. One interviewee said, “I have stopped putting work online and so have many of my friends because we are scared of our work becoming parts of these datasets with zero consent.” Another artist expressed that they do not add any alternate text or description to their images so that they are not useful for AI models. 

\textbf{Artists’ are thinking about new workflows, business models and niches.}
When asked about possible use cases of these generative AI tools, artists often tied it to their creations. For instance, the game designer said that they would use AI for generating game characters and game scenes. Another interviewee said, “Maybe it can be used in advertising or virtual reality.” Artists expressed creating many variations of one character that they had created. One interviewee said, “It’s revolutionary. It’s helped me with brainstorming. But moreover, it fascinates me what machines interpret words as. I also really like playing with words - taking tips from other creators - changing small aspects and seeing what changes in the art.” One artist on Reddit used Stable Diffusion to generate lines from their favorite book - another used AI to generate a children’s book. 

\subsection{RQ4: What are artists’ opinions about policy surrounding the development of generative AI tools?}

\textbf{Artists favor regulation around the generation and spread of AI-generated art.}
All artists we interviewed were in support of regulation around the generation or spread of AI-generated art, however, the extent to which they supported policy regulation varied. One interviewee said, “I think there should be policy regulating what data it uses to train these models. And maybe people should be declaring what is AI-generated or not while selling it for sure but I don’t think the use of the tool itself should be regulated.”, while another artist expressed support for the artists that sued big tech companies over the use of their art, and said, “I think there should also be national or global guidelines around what data can be used, and consulting people if their art has made its way in there and sort of giving them control to remove it - not sure if that is even possible. But preventing that from happening in the first place. At the very least, there should be a way for these companies to compensate these artists.”.  Even artists that were in support of generating AI art commented, “perhaps it is too nascent and perhaps it needs policing, regulations, ethical sourcing of images, royalty models, privacy and certain laws in place, but I certainly feel like I am in support of it.”.  

\textbf{Artists want rights over the usage of their art in datasets and to be fairly compensated for their work.}
Artists both on social media as well as in our interviews expressed their desire to have agency over the usage of their art, and did not support the use of their work in datasets without their consent. Further, artists want to be fairly compensated for their work. For instance, one artist appreciated the company Shutterstock’s plans to compensate artists whose work was used to train AI models. Others thought that there is no going back now that the models have been released, and policy would not help much. One interviewee said, “Also I think it’s too late now. Like Rutkowski’s art has been used so much now, that even if tomorrow there is law around it, the generated art is already out there in massive quantities. It is never going to fully disappear now - the damage there has already been done. I think in true sense, the governance and policy makers around using datasets for AI research and generative art in particular have failed us.” 

\textbf{Artists believe that AI-generated art should be marked as such.}
Artists, especially on online platforms such as Artstation strongly believed that AI-generated art should be marked as such so that it is transparent that it is not created by a person, and so that viewers can choose to not engage with it if they are opposed to AI-generated art. One interviewee said, “maybe people should be declaring what is AI-generated or not while selling it for sure but I don’t think the use of the tool itself should be regulated.” 

\textbf{Artists do not support the generation in particular artists’ styles without their consent.}
Of the concerns about the use of artists’ datasets without their consent, the most commonly cited use case was to create art in another artist’s “style” without their consent. For instance, users on Twitter were enraged when a developer who trained and shared a model to draw in the style of late artist Kim Jung Gi as homage. Another Twitter user who encountered people using generative-AI to create digital artwork resembling her style commented that she was “heartbroken” and that “everyone using AI art is dead to her.” 

\textbf{Platforms hosting art also bear responsibilit.} 
Artists recognize that policy doesn’t need to just regulate the developers and companies of these AI models, but also websites and platforms that artists use to share their artwork. When asked about their opinions about policy governing AI models to create art, one interviewee said, “I do think there should be policy for sure, and it should take into account artists’ concerns because let’s be honest, their work and data was integral to these models being even half as useful as they are today.” Another interviewee commented that “platforms should either make this scraping impossible or track how the artists’ work is being used and fairly compensate them”. Artists criticized that Artstation was removing posters of “No to AI Art” were being removed by the platform, but welcomed the platform’s decision to require creators to mark AI art as generated by AI. 

\section{Discussions}

This work is an empirical study about artists’ sentiments about AI-generated art. We learned that while artists are hopeful about possible use cases of AI in collaborative art making and fantastical generations, they are dissatisfied with the use of their artwork in AI models without their consent. It is clear that their livelihood and workflow is being impacted. They support policies regulating the usage of these data. There was variation in whether they consider AI-generated images as art at all, with some accepting it as a new medium of creation, while others not considering it human-enough or intentional-enough to be called art. Artists have also developed strategies to adapt to this new reality of AI-generated content being on their art sharing platforms. Some artists chose to unfollow those sharing AI art, while others decided not to post their art online or to not add alt text to it so that it does not get used in AI models. This reduces access to art, especially for people with visual impairments who rely on alt text to enjoy visual art. 

Art sharing platforms have also adopted different strategies to deal with the sudden proliferation of AI generated art and resulting protest asking to ban it. Platforms such as Getty Images have already banned the presence of AI Art on the website. Artstation on the other hand did not explicitly ban AI Art, but their terms and conditions stated, "Projects tagged using “NoAI” will automatically be assigned an HTML “NoAI” meta tag. This will explicitly disallow the use of the content by AI systems and mark the project so that they know they are not allowed to use it." They stated further, "ArtStation’s content guidelines do not prohibit the use of AI in the process of artwork being posted. We are soon going to require all content created by AI that is sold for commercial reuse to be explicitly labeled as created with AI, and are going to update the Terms of Service and establish a reporting process for such content that is unlabeled." DeviantArt, on the other hand, initially made all art on the platform available to AI models by default, unless artists specifically opted out. However, after protest from the users of the platform, they stated, "ALL deviations are automatically labeled as NOT authorized for use in AI datasets." DeviantArt DeviantArt also embraced AI image generation, launching a tool called DreamUp (based on Stable Diffusion) that lets anyone make pictures from text prompts. To protect artists, DreamUp is built to detect when you’re trying to ape another artist’s style, and if the artist objects, it’s supposed to stop you.

Artists are key stakeholders in the digital art space, and their concerns and goals are integral to the ethical development of these tools. Polish artist Greg Rutkowski who creates dreamy fantasy landscape, and whose name has appeared in over 10 million images and prompts generated by Stable Diffusion, expressed his concern, “It’s been just a month. What about in a year? I probably won’t be able to find my work out there because [the internet] will be flooded with AI art,” Rutkowski says. “That’s concerning.”~\cite{melissa2023}  

Furthermore, there already exists a power dynamic between technologists and artists, where artists feel left out of emerging technology that was built on top of their work, in a manner that is unethical to them. As one interviewee commented, “I think when the models are ethically sourced, artists will also be invested in adopting these tools and finding positive use cases - they truly are the main stakeholders here”. Learning from our findings, in this section we lay out recommendations for responsible and inclusive design of text-to-image AI tools geared towards art-making: 

\begin{itemize}
    \item \textbf{Ethically sourcing images: }Generative AI models should not be trained on images that are sourced without the artists’ consent. Better data organization, especially for data collected online is required for developers to scrape only ethically sourced images. 
    \item \textbf{Transparency:} Art created using AI tools should be clearly marked so, so that it can be differentiated from human-made art. 
    \item \textbf{Responsible design of sharing platforms:} Art sharing platforms should be transparent to any new users about to upload their art about how and where their data can potentially be used and the platform’s publicly about using the platform’s  data are. Platforms must adopt an opt-in model, where by default, artists' work is not available to be used by AI models unless they intentionally opt in. Different platforms may have different data sharing policies, but it should be clear to any artist using the website to upload their work. Further, art sharing sites need to adopt a clearly stated policy about the presence of AI art on their platforms. 
    \item \textbf{No consent, no style:} Development of art models to create art in a particular artists’ style, without their explicit consent or participation, should be discouraged, 
    \item \textbf{Designing with artists:} Artists can be a key future stakeholder as potential users of these tools in their workplace. Artists identified the potential of using these tools for creating imaginative fantastical content and as a collaborative tool in their own workflow. Designing the tools with artists, keeping in mind their needs and goals, could be beneficial for AI art tools. 
    \item \textbf{AI Literacy:} Our interviews and study of online discussions about AI art also reveal that artists have a very varying level of understanding about how AI-generated art works, with some artists having incorrect understanding of the algorithm. To make artists a part of the conversation, there need to be literacy efforts geared towards teaching artists about how these algorithms work, what they are capable of and how they can be used. 
    \item \textbf{Powerful tool for imagination:} Artists recognize how AI-assisted generation is helpful in imagination. Building these tools in existing brainstorming or imagination tools could be a useful method to leverage AI during the ideation process. 
\end{itemize}

\section{Conclusion}

In this work we studied artists’ hopes around the use of AI-powered art generation tools, their concerns about the proliferation of these tools, how they see these tools influencing their work, and their opinions on what policy surrounding the use of these tools must be. This work is a systematic empirical study of how visual artists perceive AI-generated art and it is clear that artists have major concerns, such as, the use of their work without their consent, the potential impact on their career and a fear of the uncertainty that these tools bring, that must be taken into consideration during the development of these tools. 

\subsection{Limitations and Future Work}

While we collected data from multiple sources such as social media platforms and artist interviews, this data does not represent the perspectives of all visual artists, and in fact serves as a very narrow view of the large world of visual artists. We collected data from artists who were already discussing AI art, hence assuming that they already have some knowledge about AI-generated art. Further, we recruited artists who post their work on social media websites. There exists a whole world of visual artists who do not have a social media presence or who we are not aware of or did not reach out to. Further, we only collected data in English and the interviews were conducted in English, thus excluding datasets from non-English speakers. Hence, while this analysis offers some insights about artists’ hopes and fears about AI-generated art, it is not a universal view of all artists, and further work is required to study the perspectives across different cultures and art forms. However, even in this small dataset, we see a wide variety of views and opinions emerge, in effect, demonstrating the complexity of the topic. These tools are evolving rapidly and so is their adoption and acceptance. As these tools become better in quality and more accessible, artist sentiment can likely shift, in both positive or negative directions. Future research to understand what has changed with the evolution of these tools with time will be an impactful contribution to the field. 

\bibliographystyle{ACM-Reference-Format}
\bibliography{sample-base}

\end{document}